# Significant reduction of lattice thermal conductivity by electron-phonon interaction in silicon with high carrier concentrations: a first-principles study

Bolin Liao[1], Bo Qiu[1], Jiawei Zhou[1], Samuel Huberman[1], Keivan Esfarjani[2,3] and Gang Chen[1*]

1. Department of Mechanical Engineering, Massachusetts Institute of Technology, Cambridge, Massachusetts, 02139, USA
2. Department of Mechanical and Aerospace Engineering, Rutgers University, Piscataway, New Jersey, 08854, USA
3. Institute for Advanced Materials, Devices and Nanotechnology, Rutgers University, Piscataway, New Jersey, 08854, USA

**Abstract**

Electron-phonon interaction has been well known to create major resistance to electron transport in metals and semiconductors, whereas less studies were directed to its effect on the phonon transport, especially in semiconductors. We calculate the phonon lifetimes due to scattering with electrons (or holes), combine them with the intrinsic lifetimes due to the anharmonic phonon-phonon interaction, all from first-principles, and evaluate the effect of the electron-phonon interaction on the lattice thermal conductivity of silicon. Unexpectedly, we find a significant reduction of the lattice thermal conductivity at room temperature as the carrier concentration goes above $10^{19}$ cm$^{-3}$ (the reduction reaches up to 45% in p-type silicon at around $10^{21}$ cm$^{-3}$), a range of great technological relevance to thermoelectric materials.

The coordinates of electrons and atomic nuclei represent the most common degrees of freedom in a solid. The full quantum mechanical treatment of the excitations in a solid thus require the solution of the Schrödinger equation involving the coordinates of all electrons and atomic nuclei, which appears intractable in most cases. A widely applied simplification, the Born-Oppenheimer approximation (BOA) [1], makes use of the fact that the electrons' mass is much smaller than that of the nuclei, and the electrons respond to the motions of the nuclei so quickly that the nuclei can be treated as static at each instant. Under BOA, the coordinates of the nuclei enter the electronic Schrödinger equation as external parameters, and in turn the electronic ground-state energy acts as part of the interaction energy between the nuclei given a specific configuration, with which the quantized excitations of the atomic nuclei, namely phonons, can be investigated separately from the electrons [2]. It is important to note, however, that BOA does not separate the electronic and atomic degrees of freedom completely, and a remaining coupling term can cause transitions between the eigenstates of the electron and phonon systems [3]. This electron-phonon interaction (EPI) problem was first studied by Bloch [4], and later understood as the main source of resistance to electrical conduction in metals and semiconductors at higher temperatures [3,5,6], and played the key role in the microscopic theory of superconductivity [7].

---

[*] Author to whom correspondence should be addressed. Electronic mail: gchen2@mit.edu

While the effect of EPI on the electron transport has been widely studied in great details and has become standard content in textbooks [3,5,6], its effect on phonon transport has received much less attention. In our opinion the reason is twofold. First of all, the carrier concentration in semiconductors for conventional microelectronic and optoelectronic applications is typically below $10^{19}$ cm$^{-3}$ [8], and as we shall show later, the impact of EPI on phonon transport in this concentration range turns out to be too small to invoke any practical interest. On the other hand, in metals with typical carrier concentration greater than $10^{22}$ cm$^{-3}$, the thermal conduction is dominated by electrons, and in most cases phonons contribute less than 10% to the total thermal conductivity [9]. Most of the existing work that were related to the effect of EPI on the lattice thermal conductivity looked into metals, pioneered by Sommerfeld and Bethe [10], and subsequently by Makinson [11] and Klemens [12]. The main conclusion is that the phonon thermal conductivity in metals is limited by EPI only at low temperatures. Early experimental attempts to measure this effect in metals were organized and reviewed by Butler and Williams [13]. The difficulty of separating the electronic and phononic thermal conductivities limited the experiments mostly to very low temperatures with high uncertainties. The classical treatment of this problem in semiconductors was provided by Ziman [3,14,15], where simplified models for the phonon dispersion, the electronic structure and the interaction matrix elements were used for a closed-form analytic formula with limited accuracy and applicability (only valid at low temperatures in degenerate semiconductors). Ensuing experiments in semiconductors also suffered from the difficulty of separating EPI from other scattering mechanisms of phonons, and thus remained qualitative and/or limited to very low temperatures [16–28]. Again the common wisdom was that the EPI would only be important on the phonon transport at low temperatures, partly due to the fact that most of the studies analyzed samples with carrier concentrations below $10^{18}$ cm$^{-3}$.

In the past two decades, the field of thermoelectrics has revived after the introduction of nanotechnology. Most of the best thermoelectric materials synthesized so far have been heavily-doped semiconductors, usually with the carrier concentration well above $10^{19}$ cm$^{-3}$ or even $10^{20}$ cm$^{-3}$ (e.g. [29] for BiSbTe, [30] for Si/Ge, [31,32] for PbTe, [33] for SnTe etc.). Moreover, a large portion of the efforts for enhancing the thermoelectric efficiency have been focused on reducing the lattice thermal conductivity via nanostructuring [32,34,35]. In this context, how the lattice thermal conductivity is affected by EPI with the carrier concentration in the range of $10^{19}$ cm$^{-3}$ to $10^{21}$ cm$^{-3}$ has become an important question to be answered in details. So far only Ziman's formula was used in modeling this effect in heavily-doped thermoelectrics [36–43], which is apparently insufficient for a modern understanding. In this Letter we attempt to answer this question accurately with calculations done fully from first-principles.

The rate of the transition of a phonon with polarization $v$, frequency $\omega$ and wavevector $\mathbf{q}$ caused by EPI can be derived in a typical structure of the Fermi's Golden Rule [3]:

$$\gamma_{\mathbf{q}v} = \frac{2\pi}{\hbar} \sum_{mn,\mathbf{k}} \left| g_{mn}^{v}(\mathbf{k},\mathbf{q}) \right|^2 \left[ f_{n\mathbf{k}}(1-f_{m\mathbf{k}+\mathbf{q}}) n_{\mathbf{q}v} \delta(\varepsilon_{m\mathbf{k}+\mathbf{q}} - \varepsilon_{n\mathbf{k}} - \omega_{\mathbf{q}v}) - f_{n\mathbf{k}}(1-f_{m\mathbf{k}-\mathbf{q}})(n_{\mathbf{q}v}+1)\delta(\varepsilon_{m\mathbf{k}-\mathbf{q}} - \varepsilon_{n\mathbf{k}} + \omega_{\mathbf{q}v}) \right], \quad (1)$$

where $g_{mn}^{v}(\mathbf{k},\mathbf{q})$ is the matrix element of one EPI process involving the given phonon and two electrons (or holes) with band indices $m$ and $n$, and wavevectors $\mathbf{k}$ and $\mathbf{k}+\mathbf{q}$ respectively, $f_{n\mathbf{k}}$ is the distribution function for electrons, $n_{\mathbf{q}v}$ is the distribution function for phonons (the superscript "0" will be used to denote the equilibrium distributions), and $\varepsilon_{n\mathbf{k}}$ is the eigen energy of an electron measured from the Fermi level. The first term in the square bracket corresponds to a phonon-absorption process while the second term a phonon-emission process. The phonon lifetime due to EPI can be defined in the following way: in equilibrium, $\gamma_{\mathbf{q}v} = 0$; if the distribution function of one phonon mode $(\mathbf{q}, v)$ is disturbed from the equilibrium by a small amount $n_{\mathbf{q}v} = n_{\mathbf{q}v}^0 + \delta n_{\mathbf{q}v}$, while assuming the electrons and other phonons are in equilibrium, the lifetime of this phonon mode $\tau_{\mathbf{q}v}^{ep}$ is defined via $\gamma_{\mathbf{q}\nu} = \frac{\delta n_{\mathbf{q}\nu}}{\tau_{\mathbf{q}\nu}^{ep}}$. This definition simplifies Eq. (1) to the expression of the phonon lifetimes:

$$\frac{1}{\tau_{\mathbf{q}\nu}^{ep}} = -\frac{2\pi}{\hbar} \sum_{mn,\mathbf{k}} \left|g_{mn}^{\nu}(\mathbf{k},\mathbf{q})\right|^2 \left(f_{n\mathbf{k}} - f_{m\mathbf{k}+\mathbf{q}}\right) \delta\left(\varepsilon_{n\mathbf{k}} - \varepsilon_{m\mathbf{k}+\mathbf{q}} - \omega_{\mathbf{q}\nu}\right). \qquad (2)$$

This expression is related to the imaginary part of the phonon self-energy $\Pi_{\mathbf{q}\nu}''$ in field-theoretical treatments of EPI: $\frac{1}{\tau_{\mathbf{q}\nu}^{ep}} = \frac{2\Pi_{\mathbf{q}\nu}''}{\hbar}$ [44]. Given that the phonon energy scale is much smaller than the electron energy scale, $f_{n\mathbf{k}} - f_{m\mathbf{k}+\mathbf{q}} \approx \frac{\partial f_{n\mathbf{k}}}{\partial \varepsilon_{n\mathbf{k}}} \hbar \omega_{\mathbf{q}\nu} = -f_{n\mathbf{k}}(1-f_{n\mathbf{k}}) \frac{\hbar \omega_{\mathbf{q}\nu}}{k_B T}$, and Eq. (2) agrees with that used by Ziman [3]. The matrix element $g_{mn}^{\nu}(\mathbf{k},\mathbf{q}) = \left(\frac{\hbar}{2m_0 \omega_{\mathbf{q}\nu}}\right)^{1/2} \langle m\mathbf{k}+\mathbf{q} | \partial_{\mathbf{q}\gamma} V | n\mathbf{k} \rangle$ [3,45], where $m_0$ is the reduced atomic mass, and $\partial_{\mathbf{q}\gamma} V$ is the variation of the electronic ground state energy with respect to a disturbance of atomic positions caused by the propagation of the phonon mode $(\mathbf{q},\nu)$. The EPI matrix elements can be calculated *ab initio* within standard density functional perturbation theory (DFPT) [46], but the phonon mesh density required for a converged EPI calculation can be rather demanding. Thanks to the recent development of an interpolation scheme using maximally-localized Wannier functions [45], EPI calculations with very fine meshes have become possible. After the EPI matrix elements are obtained, Eq. (2) can be integrated over the first Brillouin zone to generate the phonon lifetimes.

To fully evaluate the effect of EPI on the lattice thermal conductivity, the intrinsic lattice thermal conductivity limited by the phonon-phonon scattering processes must also be calculated from first-principles and used as the baseline. Several authors of this Letter have developed a first-principles framework to achieve this goal based on density functional theory (DFT) and real-space lattice dynamics [47,48]. This method has been applied to a wide range of materials and the agreements with experimental data are remarkable [49–53]. The lifetimes due to both the phonon-phonon interaction and the electron-phonon interaction are finally combined using the Mattiessen's rule [3], and the lattice thermal conductivity can be calculated as the sum of contributions from all phonon modes $\kappa = \frac{1}{3}\sum_{\mathbf{q}\nu} C_{\mathbf{q}\nu} v_{\mathbf{q}\nu}^2 \tau_{\mathbf{q}\nu}$, where $C_{\mathbf{q}\nu}$ is the mode-specific heat capacity, $v_{\mathbf{q}\nu}$ the group velocity and $\tau_{\mathbf{q}\nu}$ the total lifetime. Several authors of this paper have recently studied the thermoelectric figure of merit zT of silicon from first-principles combining the above two approaches [54].

We use the Quantum Espresso package [55] for the DFT and DFPT calculations, with a norm-conserving pseudopotential with Perdew-Burke-Ernzerhof exchange-correlation functional [56]. The EPI matrix elements are first calculated on a $12 \times 12 \times 12$ k-mesh and a $6 \times 6 \times 6$ q-mesh, and later interpolated to finer meshes using the EPW code [57]. The original code is modified to carry out the Brillouin zone integration using the tetrahedra method [58] to improve the convergence. The convergence of the phonon lifetimes due to EPI with respect to the k-mesh density is checked [59]. Results shown later are calculated on a $60 \times 60 \times 60$ k-mesh and a $18 \times 18 \times 18$ q-mesh unless otherwise stated. The details of the phonon-phonon calculation follow those in Ref. [48]. All calculations are performed at the room temperature (300K).

The scattering rates of all phonon modes due to EPI (by either electrons or holes) at the carrier concentration of $10^{21}$ cm$^{-3}$ are given in Fig. 1. Several general features can be observed. First of all, phonons near the zone center, both acoustic and optical ones, are strongly scattered by both the electrons and holes in intravalley processes. Since the phonon energy scale is much smaller than that of the electrons, phonons with larger wavevectors are less likely to be scattered by electrons, and the corresponding scattering phase space restricted by the energy and momentum selection rules is much smaller. This is reflected in the low scattering rates of phonons with intermediate wavevectors. For phonons near the zone boundary, the scattering rates due to electrons or holes are very different. In the case of scattering with electrons, the phonons near the zone boundary can efficiently participate in intervalley processes, moving electrons among the 6 equivalent pockets near the bottom of the conduction band, and the resulted scattering rates are comparable to those of the phonons near the zone center. In the case of scattering with holes, however, the intervalley processes are absent due to the sole hole-pocket, and thus the scattering rates of the phonons near the zone boundary are very low.

Since it is very difficult, if not impossible, to isolate the contributions of EPI to the lattice thermal conductivity experimentally, we are not able to directly verify our calculations via comparing with any experimental data. As a benchmark, we study the asymptotic behavior of the scattering rates of phonons near the zone center, and compare it with the existing analytic model. At the long wavelength limit, the effect of phonons on the lattice approaches a uniform strain, and thus the matrix elements $\langle m\mathbf{k}+\mathbf{q}|\partial_{\mathbf{q}\gamma}V|n\mathbf{k}\rangle$ can be replaced by a constant *deformation potential*: $D_A q$ for acoustic phonons and $D_O$ for optical phonons [5]. The presence of $q$ in the acoustic case is due to the fact that the deformation potential is proportional to the spatial derivative of the atomic displacement, while in the optical case, it is proportional to the atomic displacement *per se* [5]. With this deformation potential approximation (DPA), the asymptotic behavior of Eq. (2) can be derived without further approximations in the nondegenerate regime as [59]:

$$\frac{1}{\tau_{\mathbf{q}v}^{ep}} = \frac{(2\pi m^*)^{1/2} D_A^2 \Omega}{(k_B T)^{3/2} g_d m_0 v_s} \exp\left(-\frac{2\pi^2 m^* v_s^2}{k_B T}\right) n(E_f) \frac{\omega_{\mathbf{q}v}}{2\pi} \text{ for acoustic modes and} \quad (3)$$

$$\frac{1}{\tau_{\mathbf{q}v}^{ep}} = \left(\frac{2\pi m^*}{k_B T}\right)^{1/2} \frac{D_O^2 \Omega}{g_d m_0 \omega_O} \sinh\left(\frac{\hbar \omega_O}{2k_B T}\right) n(E_f) \exp\left(-\frac{m^* \omega_O^2}{2k_B T q^2}\right)(\hbar q)^{-1} \text{ for optical modes,} \quad (4)$$

where $m^*$ is the density-of-state effective mass of the carriers, $\Omega$ the volume of a unit cell, $g_d$ the number of equivalent carrier pockets, $v_s$ the sound velocity, $n(E_f)$ the carrier concentration with $E_f$ being the Fermi level, and $\omega_O$ the optical phonon frequency (~15 THz in silicon). Equations (3) and (4) supplement Ziman's formula in the nondegenerate regime at higher temperatures. In Fig. 2 we show the comparison between the calculated scattering rates and the analytic predictions Eqs. (3) and (4) for longitudinal acoustic (LA) and optical (LO) phonons scattered by electrons or holes (the shear strain induced by transverse phonons is a second-order effect in the DPA formalism [5] and thus does not fit in the discussion here). A $60 \times 60 \times 60$ q-mesh is used for this calculation. As predicted by Eq. (3), the scattering rates of LA modes scale linearly with the phonon frequency near the zone center, and the slope in turn linearly depends on the carrier concentration. As the carrier concentration approaches the degenerate regime, the scattering rates saturate. In the case of LO modes, the scattering rates depend on the magnitude of the wavevector in a more complex manner. Due to the anisotropy of the electron pockets, the EPI scattering rates near the zone center are more scattered compared to holes. Good agreements between the calculated scattering rates and the DPA prediction are observed with $D_A \approx 6$ eV, $D_O \approx 0.3 \times 10^8$ eV/cm for electrons and $D_A \approx 4.1$ eV, $D_O \approx 2.2 \times 10^8$ eV/cm for holes, all in a reasonable range comparing to literature [5].

Upon gaining confidence in our calculation, we proceed to compare the scattering rates of phonons due to EPI to the intrinsic phonon-phonon interactions, as shown in Fig. 3. It is clearly shown that the EPI scattering rates are at least two orders of magnitude lower than the intrinsic phonon-phonon scattering rates when the carrier concentration is below $10^{18}$ cm$^{-3}$. Above $10^{19}$ cm$^{-3}$, the EPI scattering rates start to be comparable to the intrinsic phonon-phonon scattering rates within the low-frequency region, and in fact surpass the phonon-phonon scattering rates for the low-frequency phonons when the carrier concentration reaches $10^{21}$ cm$^{-3}$. This is expected to have a major impact on the lattice thermal conductivity since most of the heat is carried by phonons with the lowest frequencies.

Figure 4 shows the calculated lattice thermal conductivity of silicon taking into account both EPI and the phonon-phonon interaction. The baseline thermal conductivity (~132 W/mK) is lower than the experimental bulk value (~145 W/mK) due to the fact that a finite q-mesh cannot capture the phonons very near the zone center that can potentially carry some heat. This problem was previously resolved using an extrapolation scheme when calculating the lattice thermal conductivity solely limited by the phonon-phonon interactions [48,49]. It is based on the assumption that the phonon scattering rates scale as $\omega^2$ near the zone center. In our case, since the scaling behaviors of the scattering rates due to EPI ($\sim \omega$) and phonon-phonon interaction ($\sim \omega^2$) are different, no straightforward extrapolation scheme is applicable. Thus we show the unextrapolated raw data here. Although the absolute value of the lattice thermal conductivity is underestimated, the relative contributions from the two scattering mechanisms should still be accurate (in fact, because the EPI scattering rates have a weaker dependence on $\omega$, we expect the relative contribution from EPI will be even higher if all the long wavelength phonon modes are considered). As expected, when the carrier concentration is below $10^{18}$ cm$^{-3}$, the effect of EPI on the lattice thermal conductivity is negligible, whereas EPI significantly reduces the lattice thermal conductivity when the carrier concentration goes above $10^{19}$ cm$^{-3}$. In particular, holes are more efficient in scattering phonons than electrons, which is probably due to the isotropic hole pockets in contrast to the anisotropic electron pockets (this finding is consistent with experimental facts where boron-doped p-type silicon has a lower thermal conductivity than phosphorous-doped n-type silicon with similar doping concentrations at the room temperature [60]), and the lattice thermal conductivity can be reduced by as much as 45% when the hole concentration reaches $10^{21}$ cm$^{-3}$.

To further analyze the effect of EPI on phonon transport, we also calculate the change of the phonon mean free paths when EPI is considered and the carrier concentration is at $10^{21}$ cm$^{-3}$. In Fig. 5 we compare the phonon mean free paths with and without EPI. Electrons and holes can efficiently scatter phonons with mean free paths longer than 100 nm, a group of phonons that carries ~70% of the total heat in silicon at 300K [48].

In summary, we carry out a first-principles calculation of the lattice thermal conductivity of silicon considering both phonon-phonon and electron-phonon interactions, and predicted a large reduction (up to 45%) of the lattice thermal conductivity due to the electron-phonon interaction at the room temperature, previously overlooked in most cases. This finding not only fills the gap of understanding of how EPI affects the lattice thermal conductivity in semiconductors when the carrier concentration is in the range of $10^{19}$ cm$^{-3}$ to $10^{21}$ cm$^{-3}$, but also has profound technological impact on the field of thermoelectrics. Although higher carrier concentration also means higher electronic thermal conductivity, it is in general much smaller than the reduction of the lattice thermal conductivity in the considered range of carrier concentrations (usually on the order of a few W/mK).

We would like to thank Sangyeop Lee, Jonathan Mendoza and Vazrik Chiloyan for helpful discussions. This article is based upon work supported partially by S$^3$TEC, an Energy Frontier Research Center funded by the U.S. Department of Energy, Office of Basic Energy Sciences, under Award No. DE-FG02-09ER46577, and partially by the Air Force Office of Scientific Research Multidisciplinary Research Program of the University Research Initiative (AFOSR MURI) via Ohio State University.

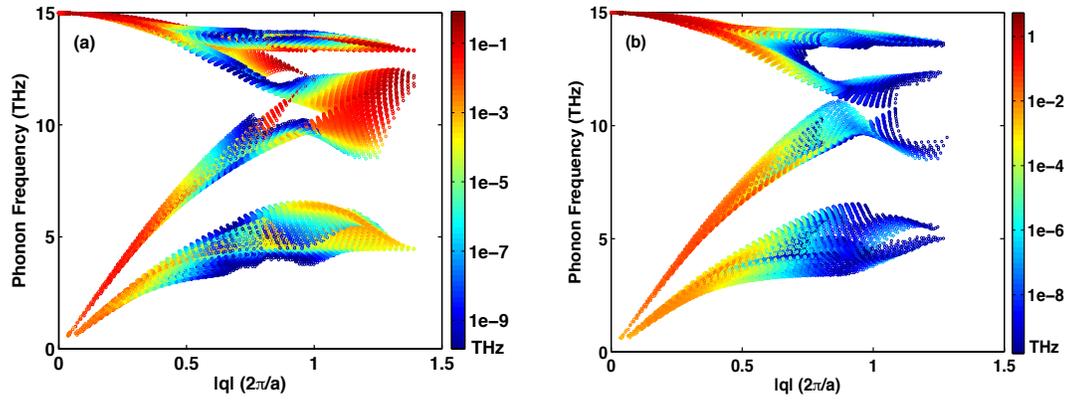

Figure 1. The scattering rates of phonons in silicon due to EPI by (a) electrons and (b) holes. The carrier concentration is $10^{21}$ cm$^{-3}$. The color denotes the scattering rates, and the white region indicates either there is no phonon mode, or the scattering rates are below the threshold rate of the calculation.

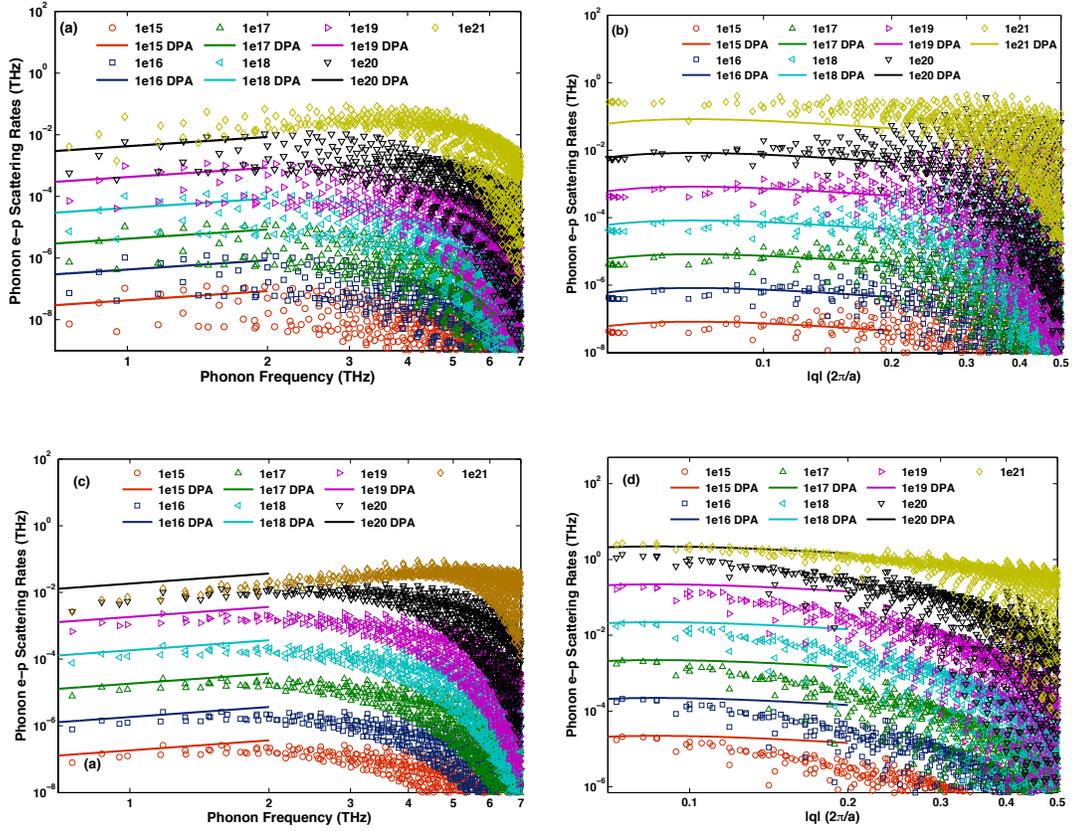

Figure 2. The asymptotic behaviors (lines) of the phonon scattering rates due to EPI, calculated from DPA, are compared with data obtained from first-principles (dots) for (a) LA modes and (b) LO modes scattered by electrons and (c) LA modes and (d) LO modes scattered by holes. A $60 \times 60 \times 60$ q-mesh is used in this calculation.

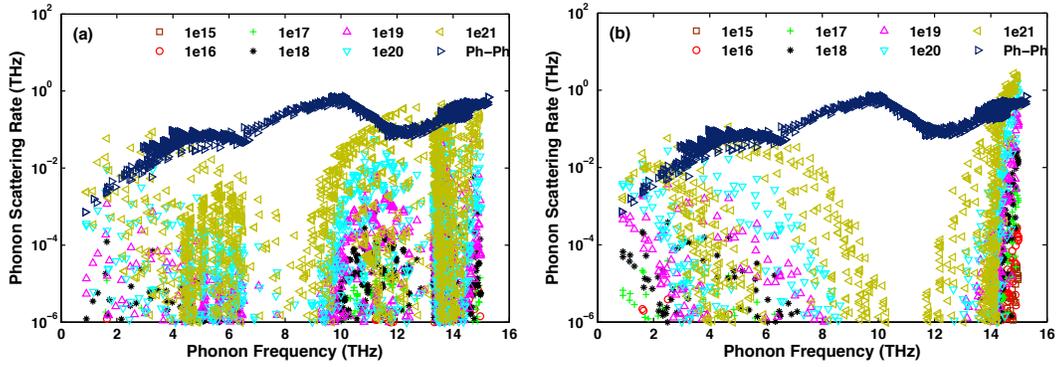

Figure 3. The phonon scattering rates due to EPI with (a) electrons and (b) holes at different carrier concentrations and the intrinsic phonon-phonon interaction. This calculation is carried out on a $18 \times 18 \times 18$ q-mesh, mainly limited by the phonon-phonon interaction calculation.

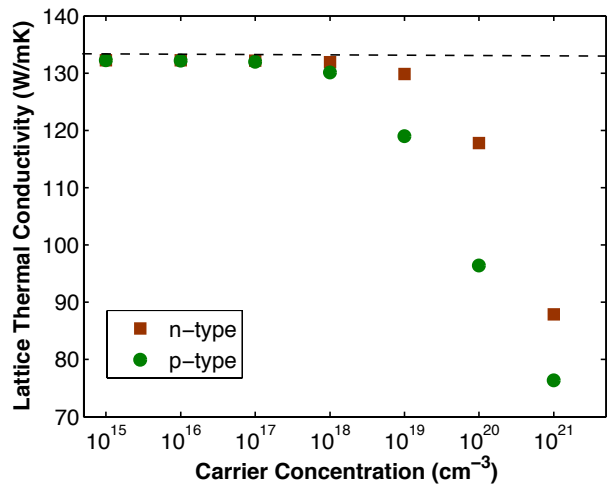

Figure 4. The lattice thermal conductivity versus the carrier concentration, taking into account both EPI and the phonon-phonon interaction.

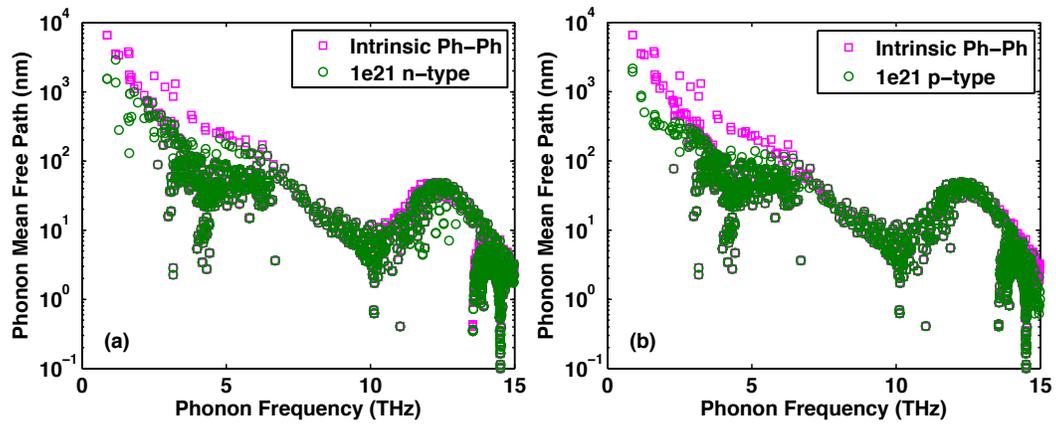

Figure 5. The phonon mean free paths with and without EPI: (a) phonons scattered by electron and (b) phonons scattered by holes. The carrier concentration is $10^{21}$ cm$^{-3}$ in both cases.